\documentclass[twocolumn,showpacs,preprintnumbers,amsmath,prc,amssymb]{revtex4-1}

\usepackage{graphicx}
\usepackage{dcolumn}
\usepackage{bm}
\usepackage[T1]{fontenc} 
\usepackage{color}

\begin{document}

\preprint{}

\title{Equation of state for neutron stars with hyperons by the variational method}

\author{H.~Togashi$^{1, 2}$}
\email{hajime.togashi@riken.jp}
\author{E.~Hiyama$^1$}
\author{Y.~Yamamoto$^1$}
\author{M.~Takano$^{2, 3}$}

\affiliation{$^{1}$RIKEN Nishina Center, RIKEN, Saitama 351-0198, Japan \\
$^{2}$Research Institute for Science and Engineering, Waseda University, Tokyo 169-8555, Japan \\
$^{3}$Department of Pure and Applied Physics, Waseda University, Tokyo 169-8555, Japan} 

\date{\today}

\begin{abstract}
We investigate the effects of the odd-state part of bare $\Lambda \Lambda$ interactions on the structure of neutron stars (NSs) by constructing equations of state (EOSs) for uniform nuclear matter containing $\Lambda$ and $\Sigma^-$ hyperons with use of the cluster variational method. 
The isoscalar part of the Argonne v18 two-nucleon potential and the Urbana IX three-nucleon potential are employed as the interactions between nucleons, whereas, as the bare $\Lambda N$ and even-state $\Lambda  \Lambda$ interactions, two-body central potentials that are determined so as to reproduce the experimental data on single- and double-$\Lambda$ hypernuclei are adopted.  
In addition, the $\Sigma^- N$  interaction is constructed so as to reproduce the empirical single-particle potential of $\Sigma^-$ in symmetric nuclear matter. 
Since the odd-state part of the $\Lambda \Lambda$ interaction is not known owing to lack of experimental data, we construct four EOSs of hyperonic nuclear matter, each with a different odd-state part of the $\Lambda \Lambda$ interaction.  
The EOS obtained for NS matter becomes stiffer as the odd-state $\Lambda \Lambda$ interaction becomes more repulsive, and correspondingly the maximum mass of NSs increases.  
It is interesting that the onset density of $\Sigma^-$ depends strongly on the repulsion of the odd-state $\Lambda \Lambda$ interaction. 
Furthermore, we take into account the three-baryon repulsive force to obtain results that are consistent with observational data on heavy NSs.  
\end{abstract}


\maketitle

\section {Introduction}
Baryon-baryon interactions are the most important ingredients for understanding the properties of hypernuclei and neutron stars (NSs). 
The nucleon-nucleon ($NN$) interaction has been extensively studied, and sophisticated $NN$ potential models have been constructed to fit the experimental $NN$ scattering data as well as the deuteron binding energy \cite{AV18, Paris, Bonn, Nijmegen}. 
In contrast, the uncertainty in hyperon-nucleon ($YN$) and hyperon-hyperon ($YY$) interactions is much larger because there exists only a small number of $YN$ scattering data and no $YY$ scattering data. 
In order to obtain informations on $YN$ and $YY$ interactions, therefore, it is necessary to study structures of hypernuclei with reliable many-body calculations. 
For example, in Refs.~\cite{Hiyama1, Hiyama2, Hiyama3}, two of the present authors (E.~H. and Y.~Y.) constructed a spin-parity-dependent $\Lambda N$ interaction so as to reproduce the experimental binding energies of light $\Lambda$ hypernuclei with the Gaussian expansion method. 
Furthermore, in Ref.~\cite{Hiyama1}, an even-state part of the $\Lambda \Lambda$ interaction is constructed so as to reproduce the experimental value of the double-$\Lambda$ binding energy extracted from the data of $^{\ \ 6}_{\Lambda \Lambda}$He (NAGARA event) \cite{NAGARA}. 

Hyperon interactions also play a crucial role in the structure of NSs. 
It has been considered that the equation of state (EOS) for dense nuclear matter becomes softer due to hyperon mixing, and the maximum mass of NSs tends to be lower than the observed masses of heavy NSs \cite{2Msolar1,2Msolar2}. 
Such hyperon mixing in NSs has been studied with various nuclear theories, such as relativistic mean field theories \cite{RMF1, RMF2, RMF3, RMF4}, Hartree-Fock approximation \cite{HF1, HF2}, quark mean field model \cite{QMF}, quantum hadrodynamics \cite{QHD}, density functional theory \cite{DFT}, and the Brueckner-Hartree-Fock theory \cite{BHF1, BHF2, BHF6, BHF3, BHF4, MPP2}.  
In particular, the Brueckner-Hartree-Fock microscopic many-body theory enables us to study the structure of NSs in terms of the bare baryon interactions.

The variational method is another powerful many-body theory for nuclear matter based on the bare nucleon interactions.  
For example, starting from the Argonne V18 (AV18) two-nucleon potential \cite{AV18} and the Urbana IX (UIX) three-nucleon potential \cite{UIX1, UIX2}, Akmal et al. (APR) performed the sophisticated Fermi Hypernetted Chain (FHNC) variational calculations to obtain energies per nucleon of pure neutron matter and symmetric nuclear matter \cite{APR}, which have been referred to as one of the standard nuclear EOSs.  
However, similar variational calculations for asymmetric nuclear matter with arbitrary proton fractions are difficult to perform.  
Furthermore, to study the EOS of hyperonic nuclear matter, the difference between the nucleon mass and hyperon masses should be taken into account.  
Consequently, only few studies use variational many-body calculations to investigate hyperonic NSs. 
A well-known example is the study by Bethe and Johnson~\cite{BJ}, who use simplified interparticle interactions because of the lack of informations at that time on hyperon interactions.  

Recently, the auxiliary field diffusion Monte Carlo (AFDMC) method was applied to hyperonic nuclear matter \cite{AFDMC3}, stressing the necessity of more constraints on the hyperon-neutron interaction.  
In principle, the AFDMC method allows us to calculate the exact energy of quantum systems. 
However, the energy calculation for arbitrary particle fractions with this method is difficult to perform as in the case with the FHNC method.  
In fact, in the study of Ref.~\cite{AFDMC3}, the energy of hyperneutron matter composed only of neutrons and $\Lambda$ hyperons are calculated for discretized sets of densities $n_{\mathrm{B}}$ and $\Lambda$-hyperon fractions $x$, and then the energy for other values of $n_{\mathrm{B}}$ and $x$ are obtained by interpolation.  

Contrary to these sophisticated variational methods, two of the present authors (H.~T. and M.~T.) recently developed a relatively simple cluster variational method for uniform nuclear matter with arbitrary proton fractions, in order to construct a microscopic nuclear EOS applicable to numerical simulations of core-collapse supernovae (SNe) \cite{K1, K2, HT1, HT2}. 
In this project, we started from the realistic nuclear Hamiltonian composed of the AV18 two-body potential and UIX three-body potential, and calculated (free) energies and other thermodynamic quantities of cold and hot asymmetric nuclear matter: 
The energies per nucleon obtained for symmetric nuclear matter and pure neutron matter at zero temperature are in good agreement with the results by APR, and, as reported in Ref.~\cite{HT1}, the mass-radius relation for NSs calculated with our EOS is consistent with observational data given in Ref.~\cite{Steiner}. 
This SN-EOS project is now in its final stage; with a Thomas-Fermi calculation we are constructing the nuclear EOS of non-uniform matter.
The results of this project will be reported in the near future.  

In the present study, we extend this reliable cluster variational method for asymmetric nuclear matter to calculate energies of hyperonic nuclear matter.  
In particular, as the first step of this extension, we take into account mixing of $\Lambda$ and $\Sigma^-$ hyperons in nuclear matter. 
Following the previous studies, we employ the AV18 and UIX potentials as realistic two- and three-nucleon potentials. 
For $\Lambda N$ and the even-state part of the $\Lambda \Lambda$ interactions, we employ two-body central potentials constructed by E. H.: The reliability of the potentials is assured in terms of the {\it ab initio} variational calculations for single- and double-$\Lambda$ hypernuclei~\cite{Hiyama1, Hiyama2, Hiyama3}. 
We also construct a new $\Sigma^- N$ potential, which reproduces the empirical single-particle potential of $\Sigma^-$ in symmetric nuclear matter at the saturation density. 
Furthermore, we need the odd-state part of the $\Lambda \Lambda$ interaction to construct the reliable EOS of hyperonic nuclear matter. 
Since the presently available experimental data on hypernuclei give no information on the odd-state part of the $\Lambda \Lambda$ interaction, we do not fix this part of the $\Lambda \Lambda$ interaction in this study; instead, we construct four models for it, and use these models to study how uncertainty in the odd-state $\Lambda \Lambda$ interaction affects NS structure. 

This paper is organized as follows. 
In Sec.~II, we treat hyperon mixing by extending the cluster variational method for asymmetric nuclear matter.  
In particular, in Sec.~II A, we introduce the Hamiltonian which is composed of bare baryon forces. 
With this Hamiltonian, we calculate in Sec.~II B the energy of hyperonic nuclear matter by the cluster variational method. 
In Sec.~III, we apply the obtained EOSs to the calculations of NS structure and discuss the effects of the uncertainty of the odd-state part of the $\Lambda \Lambda$ interaction on the properties of NSs. 
Furthermore, we examine the effect of three-baryon forces (TBFs) on the NS structure by including a three-baryon potential in our theory.  
Finally, conclusions are given in Sec.~IV. 

\section{Formalism}
\subsection{Hamiltonian}
In this section, we calculate the energy per baryon of hyperonic nuclear matter using the cluster variational method.  
For this purpose, we extend the cluster variational method for asymmetric nuclear matter reported in Ref.~\cite{HT1}; hereafter, we refer to this paper as paper I.  
In paper I, the nuclear Hamiltonian is decomposed into the two-body and the three-body parts, and the expectation value of the two-body Hamiltonian is calculated carefully so as to reproduce the results of more-sophisticated FHNC many-body calculations by APR \cite{APR}. 
Following this procedure, in this study, we first decompose the Hamiltonian $H$ of hyperonic nuclear matter into the two-body Hamiltonian $H_2$ and the three-body Hamiltonian $H_3$.  

The two-body Hamiltonian $H_2$ is written as 
\begin{equation}
H_2 = -{\textstyle\sum\limits_i} \frac{\hbar^2}{2m_i}\nabla_i^2 + {\textstyle\sum\limits_{i<j}}V_{ij}, 
\label{eq:H2}
\end{equation}
where $m_i$ is the mass of the $i$-th particle and $V_{ij}$ is the two-body potential composed of the $NN$, $YN$, and $YY$ potentials.  
As in paper I, we employ the isoscalar part of the AV18 potential \cite{AV18} as the $NN$ interaction $V_{ij}^{NN}$, i.e.,
\begin{eqnarray}
 V_{ij}^{NN} &=& {\textstyle\sum\limits_{p={\mathrm{+}}}^{\mathrm{-}}}{\textstyle\sum\limits_{s=0}^1}
        \big[V_{\mathrm{C}ps}(r_{ij}) + sV_{\mathrm{T}p}(r_{ij})S_{\mathrm{T}ij}   \nonumber   \\
&&+ sV_{\mathrm{SO}p}(r_{ij})(\mbox{\boldmath$L$}_{ij}\cdot\mbox{\boldmath$s$}) 
+ V_{\mathrm{qL}ps}\left| \mbox{\boldmath$L$}_{ij} \right| ^2   \nonumber   \\
&&+ sV_{\mathrm{qSO}p}(r_{ij})(\mbox{\boldmath$L$}_{ij}\cdot\mbox{\boldmath$s$})^2\big]P_{psij}^{\mu=NN}.
\end{eqnarray}
On the right-hand side of this equation, $p$ and $s$ are the two-nucleon relative parity and total spin, respectively; $p$ = "$+$" or "$-$" represents the even- or odd-parity state.   
$S_{\mathrm{T}ij}$ is the tensor operator, $\mbox{\boldmath$L$}_{ij}$ is the relative orbital angular momentum operator, and $P_{psij}^{\mu=NN}$ is the projection operator projecting the ($i, j$) baryon pair state on two-nucleon ($NN$) states with the relative parity $p$ and total spin $s$;  
$\mu$ represents the species of the ($i, j$) baryon pair.  

For the $\Lambda N$ interaction, we employ the single-channel interaction \cite{Hiyama2, Hiyama3} simulating the basic features of NSC97f \cite{NSC97f} expressed as 
\begin{equation}
V_{ij}^{\Lambda N} = {\textstyle\sum\limits_{p}}{\textstyle\sum\limits_{s = 0}^1}
         V_{\mathrm{C}ps}^{\mu=\Lambda N}(r_{ij}) P_{psij}^{\mu=\Lambda N}. 
 \end{equation}
Here, the $\Lambda N$-$\Sigma N$ coupling effects are renormalized into $\Lambda N$-$\Lambda N$ parts, that is, we use the central three-range Gaussian potential so as to reproduce the $\Lambda N$ scattering phase shifts calculated from the NSC97f, and then their second-range strengths of the even-state part of this potential are tuned so as to reproduce the observed energies of $0^+$ and $1^+$ spin-doublet states in $^4_{\Lambda}$H in the $NNN\Lambda$ four-body calculation. 
Furthermore, second-range strengths of the odd-state part are adjusted to reproduce the experimental values of the splitting energies of $^7_{\Lambda}$Li, as reported in Ref.~\cite{Hiyama2}. 
The explicit expression of this potential is given in Eq.~(10) of Ref.~\cite{Hiyama2}, and, in this study, we use the values in parentheses shown in Table I of Ref.~\cite{Hiyama2}. 

Contrary to the case of the $\Lambda N$ interaction, much fewer experimental data are available for the $\Sigma^- N$ interaction. 
Therefore, we construct a $\Sigma^- N$ single-channel three-range Gaussian potential simulating the radial form of the latest version of the Nijmegen model ESC08c \cite{ESC081, ESC082}.  
Then, the strength is tuned so that the single-particle potential of $\Sigma^-$ in symmetric nuclear matter is consistent with the empirical value.
Our obtained $\Sigma^- N$ interaction is noted to be of more repulsive nature than the corresponding part in ESC08c \cite{ESC082}.
The explicit expression of this $\Sigma^- N$ potential is given as follows: 
\begin{equation}
V^{\Sigma^-N}_{ij} = {\textstyle\sum\limits_{t, p, s}}{\textstyle\sum\limits_{k=1}^3} v_{k}^{(tps)} e^{-\beta_{k}r_{ij}^2}P_{psij}^{t},   
\label{eq:VSN}
\end{equation}
where $\beta_k$ are the size parameters and $v_{k}^{(tps)}$ are the strength parameters, which depend on the two-body total isospin $t$, spin $s$, and parity $p$. 
Furthermore, $P_{psij}^{t}$ in Eq.~(\ref{eq:VSN}) is the projection operator projecting a $\Sigma^- N$ pair state onto the eigenstates with respect to $p$,  $s$, and $t$.  
The values of the parameters used in Eq.~(\ref{eq:VSN}) are listed in Table~\ref{table:SN}. 
It should be noted that the $\Sigma^- N$ potential in Eq.~(\ref{eq:VSN}) is defined on the isospin basis as in the ESC08 model, whereas the particle-basis $\Sigma^- \mathrm{n}$ and $\Sigma^- \mathrm{p}$ potentials are used in the present calculations, the latter being easily obtained from the former. 
The single-particle potential of $\Sigma^-$ in symmetric nuclear matter calculated with this $\Sigma^- N$ potential is consistent with the empirical value, as discussed below.  

\begin{table}[b]
\caption{\label{table:SN}
Parameter values for the $\Sigma^- N$ potential given in Eq.~(\ref{eq:VSN}). 
$\beta_{k}$ is in fm$^{-2}$ and $v_{k}^{(tps)}$ are in MeV. 
}
\begin{ruledtabular}
\begin{tabular}{ccccccc}
\textrm{$k$}&&&&
\textrm{1}&
 \multicolumn{1}{c}{\textrm{2}}&
 \multicolumn{1}{c}{\textrm{3}}\\
\colrule
$\beta_{k}$ &&&&
 $0.250$ & $1.563$ & $8.163$  \\
\colrule
 & $t$ & $p$ & $s$ & & & \\
 & $3/2$ & $+$ & $1$ & $1.245$ & $25.26$ & $5757$  \\
 & $3/2$ & $+$ & $0$ & $-7.111$ & $-409.6$ & $8477$  \\
$v_{k}^{(tps)}$ & $3/2$ & $-$ & $1$ & $0.9283$ & $10.50$ & $4688$  \\
 & $3/2$ & $-$ & $0$ & $-9.052$ & $-182.8$ & $4390$  \\
 & $1/2$ & $+$ & $1$ & $-5.458$ & $-337.8$ & $3666$  \\
 & $1/2$ & $+$ & $0$ & $8.240$ & $340.1$ & $4799$  \\
 & $1/2$ & $-$ & $1$ & $-6.261$ & $-211.7$ & $5418$  \\
 & $1/2$ & $-$ & $0$ & $12.63$ & $-136.3$ & $24110$  \\
\end{tabular}
\end{ruledtabular}
\end{table}

For the $YY$ interactions, we only consider the $\Lambda \Lambda$ interaction, because the other $YY$ interactions cannot be determined by the experimental data on hypernuclei.  
For the $\Lambda \Lambda$ interaction, we employ the three-range Gaussian potential constructed by one of the present authors (E.~H.) and the collaborators~\cite{Hiyama1} as in the case of the $\Lambda N$ interaction.  
The even-state part of the $\Lambda \Lambda$ interaction is expressed as follows: 

\begin{equation}
V_{ij}^{\Lambda \Lambda, \mathrm{even}} = {\textstyle\sum\limits_{k=1}^3}(v_{k}^{\mathrm{even}} +v_{k}^{\sigma, \mathrm{even}}\mbox{\boldmath$\sigma$}_i\cdot\mbox{\boldmath$\sigma$}_j)e^{-\beta_{k}^{\mathrm{even}}r_{ij}^2}.  
\label{eq:VLN}
\end{equation}
Here, the values of $v_{k}^{\mathrm{even}}$, $v_{k}^{\sigma, \mathrm{even}}$ and $\beta_{k}^{\mathrm{even}}$, which are given in Table IV of Ref.~\cite{Hiyama1}, are chosen so as to reproduce the Nijmegen model F potential \cite{NSC97f, NF1, NF2} and are subsequently retuned to reproduce the experimental $\Lambda \Lambda$ binding energy given by the NAGARA event \cite{NAGARA}. 

Contrary to the even-state part, no experimental data are available to determine the odd-state part of the $\Lambda \Lambda$ interaction, because two $\Lambda$s in the experimentally known double $\Lambda$ hypernuclei are in the relative $s$ orbit. 
Therefore, in this study, we investigate how uncertainty in the odd-state part of the $\Lambda \Lambda$ interaction affects NS structure. 
For this purpose, we prepare four different models (Types 1-4) for the odd-state part of the $\Lambda \Lambda$ potential expressed as in the case of the even-state part: 
\begin{equation}
V_{ij}^{\Lambda \Lambda, \mathrm{odd}} = {\textstyle\sum\limits_{k=1}^3}(v^{\mathrm{odd}}_{k} +v_{k}^{\sigma, \mathrm{odd}}\mbox{\boldmath$\sigma$}_i\cdot\mbox{\boldmath$\sigma$}_j)e^{-\beta^{\mathrm{odd}}_{k}r_{ij}^2}.  
\label{eq:VLLodd}
\end{equation}
Here, as in the case of the $YN$ interactions, the parameters $\beta^{\mathrm{odd}}_{k}$ are chosen to be the same as $\beta_{k}^{\mathrm{even}}$ for the even-state $\Lambda \Lambda$ interaction (i.e., $\beta^{\mathrm{odd}}_1 =0.555$ fm$^{-2}$, $\beta^{\mathrm{odd}}_2 =1.656$ fm$^{-2}$, and $\beta^{\mathrm{odd}}_3 =8.163$ fm$^{-2}$ \cite{Hiyama1}). 
Furthermore, $v^{\mathrm{odd}}_3$ and $v_3^{\sigma, \mathrm{odd}}$ are chosen to be the same for all four models of the odd-state $\Lambda \Lambda$ interaction (i.e., $v^{\mathrm{odd}}_3 = 4884$ MeV and $v_3^{\sigma, \mathrm{odd}} = 915.8$ MeV).  
This implies that the repulsive core of the four models have similar strengths and ranges.  
Finally, the remaining $v^{\mathrm{odd}}_k$ and $v_k^{\sigma, \mathrm{odd}}$ ($k=1, 2$) are chosen so that the odd-state $\Lambda \Lambda$ interaction becomes monotonically more repulsive in going from Type 1 to Type 4.
As a measure of the character and strength of the odd-state interaction with a potential $V(r)$ , we employ the '$p$-wave' volume $J_{p\mathrm{-wave}}$ defined as~\cite{Hiyama4}
\begin{equation}
J_{p\mathrm{-wave}} = \int V(r) r^2 d\mbox{\boldmath$r$}.   
\end{equation}
The strength of the most attractive Type 1 interaction is chosen to be comparable to that of the odd-state of the $\Lambda N$ interaction of the Nijmegen hard core model, which is given in Ref.~\cite{Hiyama4}. 
Type 2 is chosen to be less attractive, whereas Type 3 is chosen to be slightly repulsive. 
Finally, Type 4 is the most repulsive; its strength is comparable to that of the spin-independent part of the odd-state $\Lambda N$ interaction used in this study 
(The $p$-wave volume of the spin-independent part of the odd-state $\Lambda N$ interaction is $J_{p\mathrm{-wave}}$ = +432 MeVfm$^5$).
The explicit values of $v^{\mathrm{odd}}_k$ and $v_k^{\sigma, \mathrm{odd}}$ ($k=1, 2$) are shown in Table \ref{table:LL}: 
The corresponding values of $J_{p\mathrm{-wave}}$ are also shown.  

\begin{table}[tb]
\caption{\label{table:LL}
Parameter values for the odd-state part of the $\Lambda \Lambda$ interaction, given in Eq.~(\ref{eq:VLN}), and the $p$-wave volume $J_{p\mathrm{-wave}}$. 
Values of $v_1^{\mathrm{odd}}$, $v_2^{\mathrm{odd}}$, $v_1^{\sigma, \mathrm{odd}}$, and $v_2^{\sigma, \mathrm{odd}}$ are in MeV, whereas '$p$-wave' volume $J_{p\mathrm{-wave}}$ is given in MeV fm$^{5}$. }
\begin{ruledtabular}
\begin{tabular}{ccccc}
\textrm{}&
\textrm{Type 1}&
\textrm{Type 2}&
\textrm{Type 3}&
\textrm{Type 4}\\
\colrule
$v_{1}^{\mathrm{odd}}$ & $-10.67$ & $-6.668$ & $-2.667$ & $-1.067$ \\
$v_{2}^{\mathrm{odd}}$ & $-93.51$ & $-58.44$ & $-23.37$ & $109.4$ \\
$v_{1}^{\sigma, \mathrm{odd}}$ & $0.0966$ & $0.0603$ & $0.0241$ & $0.00966$ \\
$v_{2}^{\sigma, \mathrm{odd}}$ & $16.08$ & $10.05$ & $4.020$ & $-18.81$ \\
$J_{\mathrm{p-wave}}$ & $-313$ & $-100$ & $+112$ & $+430$ \\
\end{tabular}
\end{ruledtabular}
\end{table}

In the next step, we introduce three-body interactions. 
For the nucleon sector, three-body Hamiltonian $H_3$ is expressed with the UIX three-nucleon potential $V_{ijk}$ \cite{UIX1, UIX2} as in paper I: 
\begin{equation}
H_3={\textstyle\sum\limits_{i<j<k}}V_{ijk}. 
\label{eq:H3}
\end{equation}
In this paper, we first take into account only this three-nucleon interaction. 

As will be reported in more detail later, the maximum mass of NSs with the nuclear EOS including only this three-nucleon force is smaller than the recent observational data on heavy NSs \cite{2Msolar1,2Msolar2}.  
At the last part of this paper, therefore, we will also take into account the three-body force including hyperons ($YNN$, $YYN$, and $YYY$) so as to reconcile our EOS with those observational data. 
For these hyperon sectors, we adopt a phenomenological three-body interaction which is expressed as a density dependent two-body effective potential reported in Refs.~\cite{MPP1, MPP2}. 
This effective potential includes the repulsive and attractive components, and the explicit expressions are given in Eqs.~(1) and (4) of Ref.~\cite{MPP2}. 
In this paper, we use the MPc-type parameter set in TABLE I of Ref.~\cite{MPP2} for the repulsive component. 
Here we note that, in Ref.~\cite{MPP2}, the Nijmegen extended soft core models are employed as the $YN$ and $YY$ two-baryon interactions; they are different from the present two-baryon interactions. 
Therefore, we readjust the values of the parameters in the attractive part of the TBF, $V_0$ and $\eta$ in Eq.~(4) of Ref.~\cite{MPP2}, 
so that the single-particle energy spectra of $\Lambda$ hypernuclei ($^{13}_{\ \Lambda}$C, $^{28}_{\ \Lambda}$Si, $^{51}_{\ \Lambda}$V, $^{89}_{\ \Lambda}$Y, $^{139}_{\ \ \Lambda}$La, $^{208}_{\ \ \Lambda}$Pb) 
calculated with the present two-baryon interaction (expressed as the $G$-matrix) and the present TBF reproduce their experimental values. (For the detailed procedure, see Ref.~\cite{MPP2}.) 
The readjusted values are $V_0 = -34.0$ MeV and $\eta = 7.3$ fm$^3$. 
It is noted that, when we calculate the single-particle potential of $\Lambda$ in symmetric nuclear matter with these readjusted $V_0$ and $\eta$ by the cluster variational method, the result is very close to that obtained with the $G$-matrix calculation.

\subsection{Cluster variational method for hyperonic nuclear matter}
Using the Hamiltonian composed of the bare baryon interactions explained above, we calculate the energy of hyperonic nuclear matter with the cluster variational method.  
As in paper I, we first calculate the expectation value of $H_2$ with the following Jastrow wave function: 
\begin{equation}
\mathnormal{\Psi}=\mathrm{Sym}\left[\prod_{i<j}f_{ij}\right]\mathnormal{\Phi}_\mathrm{F}, 
\end{equation}
where $\mathnormal{\Phi}_\mathrm{F}$ is the wave function of non-interacting hyperonic matter at zero temperature, and $\mathrm{Sym}[~]$ represents the symmetrizer with respect to the order of the factors in the products. 
The function $f_{ij}$ is the two-body correlation function and is expressed as
\begin{eqnarray}
f_{ij} &=& {\textstyle\sum\limits_{\mu, p, s}}
        [f_{\mathrm{C}ps}^{\mu}(r_{ij}) + sf_{\mathrm{T}p}^{\mu}(r_{ij})S_{\mathrm{T}ij}   \nonumber   \\
 && + sf_{\mathrm{SO}p}^{\mu}(r_{ij})(\mbox{\boldmath$L$}_{ij}\cdot\mbox{\boldmath$s$})]P_{psij}^{\mu}.
\label{eq:fij}
\end{eqnarray}
Here, $s$ is the two-body total spin, $p$ is the parity, and $\mu$ represents the species of the particle pair ($i$, $j$). 
In the summation in Eq.~(\ref{eq:fij}), we implicitly impose the constraint that, for two identical particles, the two-particle states suitable for the Fermi-Dirac statistics are taken.  
For example, we use the triplet-odd ($(s, p) = (1, -)$) and singlet-even ($(s, p) = (0, +)$) states for a $\Lambda\Lambda$ pair ($\mu = \Lambda \Lambda$). 
Furthermore, $f_{\mathrm{C}ps}^{\mu}(r)$, $f_{\mathrm{T}p}^{\mu}(r)$, and $f_{\mathrm{SO}p}^{\mu}(r)$ are the state-dependent central, tensor, and spin-orbit correlation functions, respectively. 
Here, we implicitly imposed that $f_{\mathrm{T}p}^{\mu}(r)$ and $f_{\mathrm{SO}p}^{\mu}(r)$ are considered only for $NN$ pairs because the corresponding noncentral and momentum-dependent parts of the interactions appear only in $NN$ interactions. 
As a result, the twenty-six correlation functions appear as independent variational functions.

As in paper I, we calculate the expectation value of $H_2$ in the two-body cluster approximation, which is appropriate for calculations of energies of hyperonic nuclear matter with arbitrary particle fractions.  
In this approximation, the two-body energy $E_2$ at a given baryon number density $n_{\mathrm{B}}$ is expressed explicitly as
\begin{widetext}
\begin{eqnarray}
E_2 (n_{\mathrm{n}}, n_{\mathrm{p}}, n_{\Lambda}, n_{\Sigma^-}) &=& E_2^{\mathrm{N}}
+ {\textstyle\sum\limits_{Y = \Lambda, \Sigma^-}} x_{Y}\frac{3\hbar^2k_{\mathrm{F}Y}}{10m_{Y}}   \nonumber   \\
&&+ 2\pi n_{\mathrm{B}}{\textstyle\sum\limits_{\mu, p, s}}
\int_0^{\infty}\Bigg[\big[f_{\mathrm{C}ps}^{\mu}(r)\big]^2V_{\mathrm{C}ps}^{\mu}(r)  
+ \frac{\hbar^2}{2m_{\mu}}\bigg[\frac{df_{\mathrm{C}ps}^{\mu}(r)}{dr}\bigg]^2\Bigg]F_{\mathrm{F}ps}^{\mu}(r)r^2dr.
\label{eq:E2}
\end{eqnarray}
\end{widetext}
Here, $n_{\mathrm{n}}$, $n_{\mathrm{p}}$, $n_{\Lambda}$, and $n_{\Sigma^-}$ are the number densities of neutron, proton, $\Lambda$, and $\Sigma^-$, respectively. 
On the right-hand side of Eq.~(\ref{eq:E2}), the first term $E_2^{\mathrm{N}}$ represents the $NN$ contribution to $E_2$ and the nucleon one-body kinetic energy. 
The explicit form of $E_2^{\mathrm{N}}$ is shown in Eq.~(8) of paper I.  
It is noted that the correlation functions in paper I are expressed with the two-nucleon total isospin $t$ and its third component rather than with $p$ and $\mu$.  
The second term on the right-hand side of Eq.~(\ref{eq:E2}) is the one-body kinetic energy of $\Lambda$ and $\Sigma^-$ hyperons, 
and $x_Y$ ($Y$ = $\Lambda$, $\Sigma^-$) are the hyperon fractions defined by $x_Y$ = $n_Y/n_{\mathrm{B}}$.
Furthermore, $m_Y$ and $k_{\mathrm{F}Y}$ represent the rest mass of a hyperon and the Fermi wave number, respectively. 
The last term on the right-hand side of Eq.~(\ref{eq:E2}) is the sum of the potential energy and the kinetic energy induced by the $YN$ and $YY$ correlations, with $m_\mu$ and $F_{\mathrm{F}ps}^{\mu}(r)$ being given by 
\begin{equation}
m_\mu = \frac{m_{b}m_{b'}}{m_{b} + m_{b'}}, 
\label{eq:mmu}
\end{equation}

\begin{eqnarray}
F_{\mathrm{F}ps}^{\mu}(r) &=& \frac{2s+1}{4}x_{b} x_{b'}   \nonumber   \\
&& \times \bigg\{1 + \epsilon_p \bigg[3\frac{j_1(\xi_{b}r)}{\xi_{b}r}\bigg]\bigg[3\frac{j_1(\xi_{b'}r)}{\xi_{b'}r}\bigg]\bigg\}. 
\label{eq:FF}
\end{eqnarray}
Here, the subscripts ($b$, $b'$) represent the species of two baryons specified by $\mu$, e.g., ($b$, $b'$) $=$ ($\Lambda$, $\mathrm{n}$) for $\mu=\Lambda\mathrm{n}$, and $\xi_{b}$ is defined by $\xi_{b} = 2m_{\mu}k_{\mathrm{F}b}/m_b$. 
In Eq.~(\ref{eq:FF}), $\epsilon_p = +1$ or $-1$ for $p = +$ or $-$, respectively.  
The differences in masses between $N$, $\Lambda$, and $\Sigma^-$ are taken into account as the reduced masses $m_\mu$ and the corresponding $\xi_{b}$.   
Here we note that, in the calculations of $E_2^{\mathrm{N}}$ and $m_{\mu}$, the proton mass $m_\mathrm{p}$ is set equal to the neutron mass $m_\mathrm{n}$ following the approach used in paper I for asymmetric nuclear matter.  

Next, we minimize $E_2$ with respect to $f_{\mathrm{C}ps}^{\mu}(r)$, $f_{\mathrm{T}p}^{\mu}(r)$, and $f_{\mathrm{SO}p}^{\mu}(r)$ by solving the Euler-Lagrange equations derived from Eq.~(\ref{eq:E2}). 
In this minimization procedure, we impose two conditions in order to compensate the lack of the higher-order cluster terms.  
The first condition is the extended Mayer's condition, whose explicit form for $YN$ and $YY$ pairs is given as  
\begin{equation}
4\pi n_{\mathrm{B}} \int_0^{\infty} \big\{[f_{\mathrm{C}ps}^{\mu}(r)]^2-1\big\}F_{\mathrm{F}ps}^{\mu}(r)r^2dr=0, 
\end{equation}
whereas the expression for $NN$ pairs is given in Eq.~(15) of paper I. 
This condition, which implies particle-number conservation for each channel of ($\mu, p, s$), is taken into account by the Lagrange-multiplier method. 
The second condition is the healing distance condition, which implies that the correlation between two particles vanishes when the distance $r$ between those two particles is larger than the healing distance $r_{\mathrm{h}}$.  
In paper I, we imposed for asymmetric nuclear matter that $r_{\mathrm{h}}$ be proportional to the mean distance between nucleons; $r_{\mathrm{h}}=a_{\mathrm{h}}r_0$, where $r_0$ is the radius of a sphere whose volume is $1/n_{\mathrm{B}}$, and the coefficient  is chosen to be $a_{\mathrm{h}} = 1.76$ \cite{K1} so that the results obtained for $E_2$ of neutron matter and symmetric nuclear matter are consistent with the results of the FHNC calculations by APR \cite{APR}.  
As an extension of this theory to hyperonic nuclear matter, we adopt the same value $a_{\mathrm{h}}=1.76$ for $YN$ and $YY$ pairs.  

Next, we calculate the nuclear three-body energy $E_3^{\mathrm{N}}$ caused by the three-nucleon force. 
Following the method used in paper I, $E_3^{\mathrm{N}}$ is expressed as 
\begin{equation}
E_3^{\mathrm{N}} = \sum_{i=\mathrm{R},{2\pi}}\langle \alpha_i H^i_3 \rangle_{\mathrm{F}} + E_{\mathrm{corr}} \label{eq:E3}. 
\end{equation}
Here, $H^\mathrm{R}_3$ and $H^\mathrm{2\pi}_3$ are the three-body Hamiltonians composed of the repulsive and 2$\pi$-exchange components of the UIX three-nucleon potential, respectively ($H_3=H^\mathrm{R}_3+H^\mathrm{2\pi}_3$).  
The bracket with the subscript F represents the expectation value with the degenerate Fermi-gas wave function.  
The coefficients $\alpha_i$ represent the corrections with respect to correlations among nucleons and the possible relativistic boost, which are treated in the EOS by APR for symmetric nuclear matter and pure neutron matter in a more sophisticated manner.  
The additional correction term $E_{\mathrm{corr}}$ is an explicit function of $n_{\mathrm{n}}$ and $n_{\mathrm{p}}$ including two parameters; the functional form is chosen to be the same as in the EOS of APR for symmetric nuclear matter.  
As reported in paper I, $\alpha_i$ and two parameters appearing in $E_3^{\mathrm{N}}$ are tuned so that the obtained total energy per nucleon $E^{\mathrm{N}} = E_2^{\mathrm{N}} + E_3^{\mathrm{N}}$ of nuclear matter reproduces the empirical saturation density $n_0 = 0.16$ fm$^{-3}$, saturation energy $E_0 = -16.09$ MeV, incompressibility $K = 245$ MeV and symmetry energy $E_{\mathrm{sym}} = 30.0$ MeV. 
Then, the total energy per baryon of hyperonic nuclear matter $E$ is expressed as 
\begin{equation}
E (n_{\mathrm{n}}, n_{\mathrm{p}}, n_{\Lambda}, n_{\Sigma^-}) = E_2 (n_{\mathrm{n}}, n_{\mathrm{p}}, n_{\Lambda}, n_{\Sigma^-}) + E_3^{\mathrm{N}}. 
\end{equation}

\begin{figure}[t]
\begin{center}
\includegraphics[width=8cm]{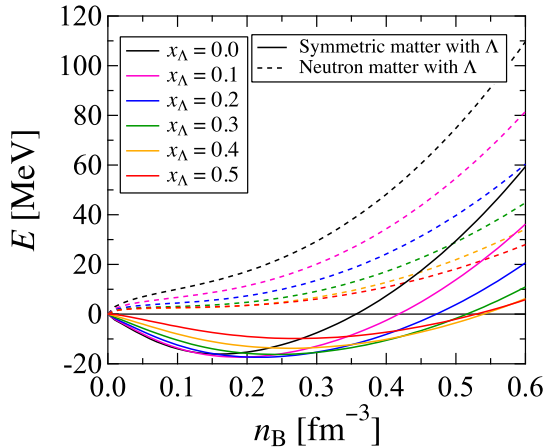}
\caption{(Color online) Energies per baryon $E$ of hyperonic nuclear matter as functions of the baryon number density $n_{\mathrm{B}}$  for various values of $\Lambda$ fractions $x_{\Lambda}$ with the most attractive odd-state part of $\Lambda \Lambda$ interaction (Type 1).  
The solid curves represent the case of $x_\mathrm{p} =x_\mathrm{n}$ while the dashed curves correspond to the case of $x_\mathrm{p}=0$.}
\label{fig:ene}
\end{center}
\end{figure}

Figure \ref{fig:ene} shows the total energies per baryon $E$ as functions of $n_{\mathrm{B}}$ with the Type 1 odd-state $\Lambda \Lambda$ interaction.  
The solid curves correspond to hyperonic nuclear matter with the proton fraction being equal to the neutron fraction ($x_\mathrm{p} = x_\mathrm{n}$); when the $\Lambda$ fraction is zero, it reduces to the result for symmetric nuclear matter, and the corresponding energy per nucleon $E$ reproduces the empirical saturation point, as  mentioned above. 
As the $\Lambda$ fraction increases, $E$ decreases at relatively high densities because $\Lambda$ hyperons occupy single-particle states with energies  much lower than those of highly degenerate nucleons.  
On the other hand, at relatively low densities, $E$ increases with the $\Lambda$ fraction because the attractive contribution from the nuclear force at these densities is stronger than that from the hyperonic interaction. 
The dashed curves in Fig. \ref{fig:ene} correspond to matter without protons; when the $\Lambda$ fraction is zero, the matter reduces to pure neutron matter.   
In this proton-less state, $E$ decreases with the $\Lambda$ fraction, similar to the case for $x_\mathrm{p} = x_\mathrm{n}$ matter at high densities.  

Given the energy $E$ calculated for hyperonic nuclear matter, we next calculate the single-particle potentials for a hyperon in nuclear matter $U_Y^0$ ($Y$ = $\Lambda$, $\Sigma^-$) (the explicit expression is given in Appendix A). 
The  $U_{\Lambda}^0$ obtained for symmetric nuclear matter at the saturation density $n_{\mathrm{B}}=n_0$ is $U_{\Lambda}^0=-43$ MeV, which is reasonably consistent with the empirical value \cite{SPL} and close to the results obtained with the $G$-matrix calculations \cite{BHF4, MPP2, NSC97f, BHF5}.   
For $\Sigma^-$ hyperons, $U_{\Sigma^-}^0 = +12$ MeV for symmetric nuclear matter at $n_{\mathrm{B}}=n_0$, which is consistent with the experimentally suggested value \cite{Gal}. 
This result is also consistent with results of $G$-matrix calculations with the ESC08 potential \cite{BHF5} and of the chiral effective field theory \cite{EFT}. 
 
Finally, the energy $E$ obtained for hyperonic nuclear matter is applied to calculations of NS structure.  
For this purpose, the total energy density $\epsilon$ of hyperonic nuclear matter including the rest mass energy of baryons is expressed as follows: 
\begin{equation}
\epsilon = {\textstyle\sum\limits_{b = \mathrm{n}, \mathrm{p}, \Lambda, \Sigma^-}} n_b m_b + n_{\mathrm{B}} E + \epsilon_e + \epsilon_{\mu},
\label{eq:epsilon}
\end{equation}
where $\epsilon_e$ and $\epsilon_{\mu}$ are energy densities of electrons and muons, respectively.  
These leptons are treated as the relativistic non-interacting Fermi gases. 
We note that in the first term on the right-hand side of Eq.~(\ref{eq:epsilon}), $m_\mathrm{p}$ is the proper proton mass, which is different from the case for Eq.~(\ref{eq:E2}).  
Then, $\epsilon$ is minimized with respect to the fractions of all the species $x_i$ ($i$ = $\mathrm{n}$, $\mathrm{p}$, $\Lambda$, $\Sigma^-$, $\mathrm{e}^-$, and $\mu^-$)  constrained by the baryon-number conservation ($n_{\mathrm{B}} = n_{\mathrm{n}} + n_{\mathrm{p}} + n_{\Lambda} + n_{\Sigma^-}$) and charge neutrality ($n_{\mathrm{p}} = n_{\Sigma^-} + n_{e^-} + n_{\mu^-}$) to obtain the energy density of NS matter $\epsilon_{\mathrm{NS}}$. 

\section{Application to neutron stars}
In this section, we investigate the effects of the odd-state part of the $\Lambda \Lambda$ interaction on the structure of NSs. 
For this purpose, as mentioned above, we calculate four EOSs of hyperonic nuclear matter using four $\Lambda \Lambda$ interactions whose odd-state parts differ from one another, as shown in Table~\ref{table:LL}. 
In particular, the repulsive effect of the odd-state $\Lambda \Lambda$ interaction increases monotonically in going from Type 1 to Type 4: 
Type 1 is the most attractive and is similar to Nijmegen hard-core model \cite{Hiyama4}.  
Type 2 is less attractive, Type 3 is slightly repulsive, and Type 4 is the most repulsive with its repulsion being comparable to the odd-state repulsion of the $\Lambda N$ interaction.  
With those $\Lambda \Lambda$ interactions, we calculate the energy per baryon $E$ for hyperonic nuclear matter using the cluster variational method. 
Finally, we calculate the EOS of NS matter as a charge-neutral, $\beta$-stable mixture of n, p, $\Lambda$, $\Sigma^-$, $e^-$, and $\mu^-$ at zero temperature. 

\begin{figure}[t]
\begin{center}
\includegraphics[width=8cm]{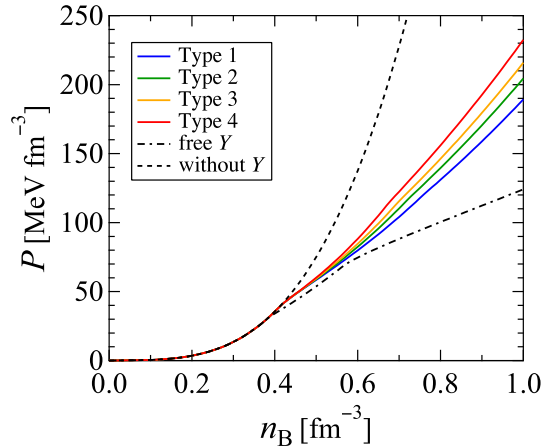}
\caption{(Color online) Pressures $P$ of NS matter with interacting hyperons by the four different odd-state parts of $\Lambda \Lambda$ interaction in Table~\ref{table:LL} as functions of the baryon number density $n_{\mathrm{B}}$. 
The pressures without hyperons (without $Y$) and with noninteracting hyperons (free $Y$) are also shown.}
\label{fig:pre}
\end{center}
\end{figure}

Figure~\ref{fig:pre} shows the pressures $P$ of NS matter derived from the energy densities of NS matter $\epsilon_{\mathrm{NS}}$ through the thermodynamic relation. 
The figure also shows the pressure of pure nucleon matter without hyperons ($x_{\Lambda}=x_{\Sigma^-}=0$) (dotted line) and that with free hyperons (dashed-dotted line).  
In the latter case, we switch off the $YN$ and $YY$ interactions. 
It is seen that the mixing of free hyperons strongly softens the EOS of NS matter at $n_{\mathrm{B}} \gtrsim 0.39$ fm$^{-3}$, as discussed below. 
The four solid lines show the pressures obtained with the above-mentioned four hyperon interactions. 
These four EOSs are softer than the EOS of pure nucleon matter.
Moreover, the figure shows that the EOS becomes stiffer as the odd-state $\Lambda \Lambda$ interaction becomes more repulsive.  
These EOSs are stiffer than those for free hyperons because, as will be discussed below, the onset density of $\Sigma^-$ with free hyperons is much lower than those with interacting hyperons.  

\begin{figure}[t]
\begin{center}
\includegraphics[width=8cm]{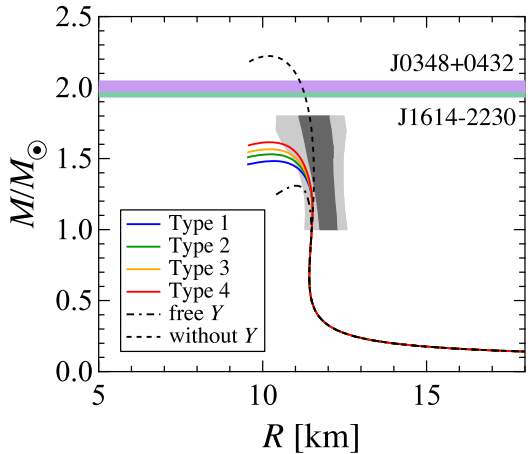}
\caption{(Color online) Mass-radius relations of NSs with the four EOSs of NS matter that correspond to the different odd-state parts of $\Lambda \Lambda$ interactions. 
The results for nuclear matter without hyperons (without $Y$) and with noninteracting hyperons (free $Y$) are also shown. 
The horizontal green and purple bands indicate the masses of PSRs J1614-2230 \cite{2Msolar1} and J 0348+0432 \cite{2Msolar2}. 
The shaded region denotes the mass-radius region suggested in Ref. \cite{Steiner}. }
\label{fig:MR1}
\end{center}
\end{figure}

With these obtained EOSs of NS matter, we solve the Tolman-Oppenheimer-Volkoff equations to obtain the mass-radius relations of NSs. 
For the NS crust region, we employ the EOS obtained with the Thomas-Fermi calculation from Ref.~\cite{K2}.  
Since the present EOS of uniform nuclear matter is used in the Thomas-Fermi calculation, this crust EOS is consistent with the present EOS for uniform hyperonic matter.  

The obtained mass-radius relations of NSs with various hyperon interactions are shown in Fig.~\ref{fig:MR1}: 
Also shown are the results for pure nucleon matter (dotted line) and with free hyperons (dashed-dotted line). 
The maximum mass of NSs with free hyperons is 1.31 $M_\odot$, which is much less than that for pure nucleon matter (2.22 $M_\odot$). 
Even with interacting hyperons (see four solid lines in Fig.~\ref{fig:MR1}), the maximum masses are less than that for pure nucleon matter.  
In other words, the hyperon mixing reduces the maximum mass of NSs because of the relative softness of hyperonic nuclear matter, as shown in Fig.~\ref{fig:pre}. 
This result is consistent with those obtained by other calculations 
such as relativistic mean field theories \cite{RMF2, RMF3, RMF4}, Hartree-Fock theories \cite{HF1, HF2} and Brueckner-Hartree-Fock theories \cite{BHF1, BHF2, BHF6, BHF3, BHF4, MPP2}. 
For example, the maximum mass of NSs calculated in the Brueckner-Hartree-Fock theory with the AV18 and UIX potentials for the nucleon sector, and the Nijmegen soft-core $YN$ and $YY$ potentials (NSC97e) for the hyperon sector is 1.31 $M_\odot$ as reported in Ref.~\cite{BHF6}. 
Figure ~\ref{fig:MR1} also shows that the maximum masses of NSs with interacting hyperons are larger than that with free hyperons.  
This tendency also appears in Fig.~\ref{fig:pre}, where all four EOSs with interacting hyperons are stiffer than the EOS with free hyperons. 

Next, we discuss how the odd-state part of the $\Lambda \Lambda$ interaction influences the maximum mass of NSs. 
Figure~\ref{fig:MR1} shows that the maximum mass of NSs increases as the strength of the odd-state $\Lambda \Lambda$ repulsion increases.  
In fact, with the most attractive odd-state $\Lambda \Lambda$ interaction (Type 1), the maximum mass of NSs is 1.48$M_\odot$, whereas, with the most repulsive odd-state $\Lambda \Lambda$ interaction (Type 4) the maximum mass is 1.62$M_\odot$; thus the maximum mass increases by about 9\% in going from Type 1 to Type 4. 
The explicit values of the maximum mass of NSs are shown in Table~\ref{tab:NS}. 

In Fig.~~\ref{fig:MR1}, the horizontal green and purple bands indicate the masses of PSRs J1614-2230 (1.97$\pm$0.04 $M_\odot$) \cite{2Msolar1} and J0348+0432 (2.01$\pm$0.04 $M_\odot$) \cite{2Msolar2}, respectively.  
In addition, the shaded region represents the observationally suggested mass-radius region analyzed in Ref.~\cite{Steiner}. 
The mass-radius relations with the present EOSs are consistent with the latter observational data.  
However, the masses of the heavy NSs can not be explained with the present EOSs, even for the most repulsive $\Lambda \Lambda$ interaction (Type 4).  
Studies with other many-body approaches also encounter this difficulty, and many trials have been made to solve this problem, one of which is to consider the three-baryon repulsive forces \cite{BHF2, BHF3, MPP2}.  
Thus, we report below on the improvement resulting from consideration of the three-baryon repulsive forces.  

Before discussing the effect of including the three-baryon repulsive force, we investigate the effects of the $\Lambda \Lambda$ interaction on the chemical composition of NS matter.  
\begin{figure}[t]
\begin{center}
\includegraphics[width=7.28cm]{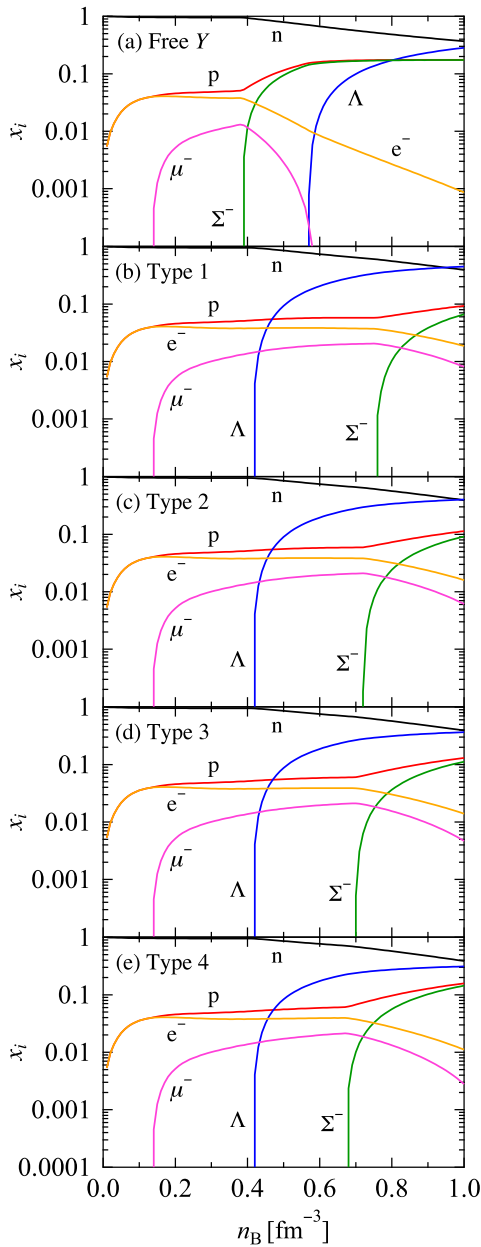}
\caption{(Color online) The fractions of particles $x_i$ of NS matter as functions of the baryon number density $n_{\mathrm{B}}$ for various odd-state parts of the $\Lambda \Lambda$ interaction; 
(a) no interaction, (b) most attractive, (c) less attractive, (d) slightly repulsive, (e) most repulsive.  }
\label{fig:frac1}
\end{center}
\end{figure}
Figure~\ref{fig:frac1} shows the fractions of neutrons, protons, $\Lambda$, $\Sigma^-$, electrons and muons $x_i$ ($i$ = n, p, $\Lambda$, $\Sigma^-$, e$^-$, and $\mu^-$) as functions of the baryon number density $n_{\mathrm{B}}$. 
Figure~\ref{fig:frac1} (a) shows the particle fractions with free hyperons. 
In this case, $\Sigma^-$ is the first hyperon to appear; its onset density is 0.39 fm$^{-3}$.  
As the $\Sigma^-$ fraction increases, the proton fraction increases, and these two fractions approach each other due to charge neutrality: 
At $n_{\mathrm{B}} \gtrsim 0.6$ fm$^{-3}$, these fractions are almost indistinguishable in this figure.
Owing to baryon number conservation, the neutron fraction decreases as the $\Sigma^-$ fraction increases.  
In contrast with the proton fraction, the fractions of leptons, which are much lighter than $\Sigma^-$, decrease with increasing $\Sigma^-$ fraction. 
The onset density of $\Lambda$ hyperons is 0.57 fm$^{-3}$, which is much higher than that of $\Sigma^-$ hyperons.  
The $\Lambda$ fraction increases with $n_{\mathrm{B}}$ and, at $n_{\mathrm{B}} \sim 1.0 $ fm$^{-3}$, the $\Lambda$ fraction becomes comparable to the neutron fraction.  
These results are consistent with those obtained with the Brueckner-Hartree-Fock calculations ~\cite{BHF1}. 

In the case with interacting hyperons shown in Fig.~\ref{fig:frac1}(b)--~\ref{fig:frac1}(e), the compositions of NS matter are quite different from that with free hyperons (Fig.~4(a)). 
In Fig.~\ref{fig:frac1} (b), NS matter is composed only of nucleons and leptons at $n_{\mathrm{B}}$ < 0.42 fm$^{-3}$.  
Contrary to the case with free hyperons (Fig.~4 (a)), the first hyperon to appear is $\Lambda$, and its onset density is 0.42 fm$^{-3}$. 
This value is lower than that with free hyperons because the repulsive $\Sigma^-$ interaction increases the onset density of $\Sigma^-$, which causes the mixing of $\Lambda$ hyperons at a relatively low density.  
As the $\Lambda$ fraction increases, the nucleon fractions decrease due to the baryon number conservation. 
The charged-lepton fractions also decrease with the proton fraction due to the charge neutrality condition. 
At $n_{\mathrm{B}} = 0.76$ fm$^{-3}$, $\Sigma^-$ hyperons appear in NS matter. 
This onset density is quite larger than that with free hyperons due to the repulsive $\Sigma^-N$ interaction.  
We note that the first hyperon to appear in this study is different from the result in the Brueckner-Hartree-Fock calculation reported in Ref. \cite{BHF6}, where $\Sigma^-$ hyperons appear at the density of about 0.35 fm$^{-3}$ before $\Lambda$ hyperons appear.  
One of the reasons is that the $\Sigma^- \mathrm{n}$ potential in the present study is rather repulsive while an attractive $\Sigma^- \mathrm{n}$ interaction is adopted in Ref.~\cite{BHF6}. 
Correspondingly, our maximum masses of NSs are slightly higher than that with the result in Ref.~\cite{BHF6}, because the onset densities of hyperons are higher in our results. 

For other cases (Types 2--4), similar tendencies are observed in Figs.~\ref{fig:frac1}(c)--\ref{fig:frac1}(e). 
In particular, the onset density of $\Lambda$ hyperons is insensitive to the odd-state $\Lambda \Lambda$ interaction because the $\Lambda \Lambda$ interaction becomes relevant in systems with many $\Lambda$ particles. 
As a result, the $\Lambda \Lambda$ interaction has a relatively large effects on the $\Lambda$ fraction in the high-density region. 
In fact, at $n_{\mathrm{B}} \gtrsim 0.70$ fm$^{-3}$, the $\Lambda$ fraction of Type 1 is larger than that of Type 4. 
Interestingly, the onset densities of the $\Sigma^-$ hyperons differ from one another: 
For the most attractive odd-state $\Lambda \Lambda$ interaction (Type 1), the onset density of $\Sigma^-$ is 0.76 fm$^{-3}$, whereas for the most repulsive odd-state $\Lambda \Lambda$ interaction (Type 4), it is 0.68 fm$^{-3}$. 
In other words, as the odd-state $\Lambda \Lambda$ interaction becomes more repulsive, the onset density of $\Sigma^-$ hyperons decreases. 
This result is attributed to the more repulsive $\Lambda \Lambda$ interaction stiffening the $\mathrm{np}\Lambda$ matter, which results in $\Sigma^-$ mixing at a lower density. 
Consequently, the odd-state $\Lambda \Lambda$ interaction strongly affects the onset density of $\Sigma^-$ hyperons rather than that of $\Lambda$ hyperons. 
The explicit values of the onset densities of $\Lambda$ and $\Sigma^-$ hyperons are shown in Table~\ref{tab:NS}. 

\begin{table*}
\caption{\label{tab:NS} The maximum masses of NSs and the onset densities of hyperons ($\Lambda, \Sigma^-$) for different hyperon interactions. 
Values of maximum masses are in the unit of $M_\odot$ and onset densities are given in fm$^{-3}$. }
\begin{ruledtabular}
\begin{tabular}{lccccccc}
\textrm{$\Lambda \Lambda$ interaction}&
\textrm{Type 1}&
\textrm{Type 2}&
\textrm{Type 3}&
\textrm{Type 4}&
\textrm{Free $Y$}&
\textrm{Without $Y$}&
\textrm{Observations}\\
\colrule
Maximum mass & 1.48 & 1.53 & 1.57 & 1.62 & 1.31 & 2.22 & 1.97 $\pm$ 0.04 \cite{2Msolar1} \\ 
& & & & & & & 2.01 $\pm$ 0.04 \cite{2Msolar2}  \\
Onset density of $\Lambda$ & 0.42 & 0.42 & 0.42 & 0.42 & 0.57 & --- & --- \\
Onset density of $\Sigma^-$ & 0.76 & 0.72 & 0.70 & 0.68 & 0.39 & --- & ---\\
\end{tabular}
\end{ruledtabular}
\end{table*}

Despite using reliable hyperon interactions to reproduce the experimental data on $\Lambda$ hypernuclei, as discussed above, the obtained maximum masses of NSs are less than the observed masses of PSRs J1614-2230 and J0348+0432.
We therefore take into account a phenomenological TBF which is expressed as a density dependent two-body effective potential as reported in Sec. IIA. 
Then, $U_{\Lambda}^0=-40$ MeV is obtained for symmetric nuclear matter at the saturation density $n_{\mathrm{B}}=n_0$, which is slightly higher than the result without TBF.
Furthermore, as in the case without the TBF, the value of $U_{\Lambda}^0$ with the TBF is also close to the result of the $G$-matrix calculation based on the ESC08 potential with the MPc-type TBF ($-37.4$ MeV) \cite{MPP2}.
Here, it is noted that, even in this case, we retain the UIX three-nucleon potential for the nucleon sector, because the EOS of nucleon matter is well established with this three-nucleon potential, as reported in paper I. 

\begin{figure}[t]
\begin{center}
\includegraphics[width=8cm]{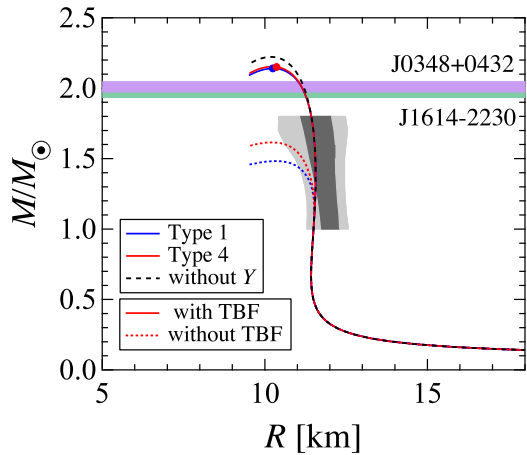}
\caption{(Color online) Mass-radius relations of NSs obtained from EOSs based on the most attractive and most repulsive odd-state part of the $\Lambda \Lambda$ interaction (Types 1 and 4, respectively) with and without phenomenological three-baryon forces (TBF). 
The filled circle represents the NS for which the central density is equal to the critical density $n_{\mathrm{c}}$. 
The result of nuclear matter without any hyperons (without $Y$) is also shown. }
 \label{fig:MR2}
\end{center}
\end{figure}

Figure~\ref{fig:MR2} shows the mass-radius relations of NSs obtained with the EOSs including TBF.  
The result with the most attractive odd-state $\Lambda \Lambda$ interaction (Type 1) and that with the most repulsive one (Type 4) are shown in this figure:  
Also shown are the results for pure nucleon matter (the black dashed curve) and for hyperon matter without the TBF (dotted curves for Types 1 and Type 4).  
The maximum masses with the TBF become larger than those without the TBF, whereas, even with the TBF, the maximum masses are less than that of pure nucleon matter. 
With the TBF, the NS structures with the Type 1 EOS is hardly distinguishable from that with the Type 4 EOS, and the maximum masses are about $2.15 M_\odot$ for both cases. 
Namely, due to the strong repulsion within the three-baryon system, we obtained results that are reasonably consistent with the observational data.  
It should be noted that, at densities higher than the critical density $n_{\mathrm{c}}$ = 1.13 fm$^{-3}$ (1.08 fm$^{-3}$) for Type 1 (Type 4), 
causality is violated in the EOSs with the TBF because the sound velocity exceeds the speed of light.
Therefore, the NS solutions with central densities being higher than $n_{\mathrm{c}}$ are not appropriate.  
However, even for densities lower than $n_{\mathrm{c}}$, the NS solutions of the EOSs are consistent with observational data, 
i.e., the NS mass at the central density of $n_{\mathrm{c}}$ is 2.14 $M_\odot$ (2.15 $M_\odot$) for the Type 1 EOS (Type 4 EOS). 

There are some other studies predicting NSs with the masses of about 2$M_{\odot}$ by introducing appropriate TBFs for hyperons~\cite{AFDMC3, MPP2}.  
On the other hand, the conclusion of Ref.~\cite{BHF3} is at variance with ours: 
In that study, various phenomenological TBFs are adopted but all the models fail to predict the 2$M_{\odot}$ NSs. 
Since the maximum masses without phenomenological TBFs for hyperons obtained in that study are close to our results, this situation implies that the maximum mass of NSs is sensitive to the details of the TBF. 
For example, we take into account the $YNN$, $YYN$, and $YYY$ interactions, while only the $YNN$ interaction is considered in Ref.~\cite{BHF3}: This fact may be a key to solve the problem.  

\begin{figure}[t]
\begin{center}
\includegraphics[width=8cm]{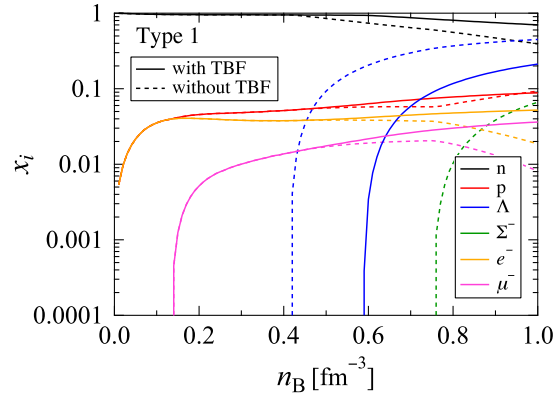}
\caption{(Color online) The fractions of particles $x_i$ of NS matter as functions of the baryon number density $n_{\mathrm{B}}$ based on the most attractive odd-state part of the $\Lambda \Lambda$ interaction (Type 1).
The results without the TBF are also shown.}
 \label{fig:frac2}
\end{center}
\end{figure}

The fractions of particles $x_i$ in NS matter with the TBF for the most attractive odd-state $\Lambda \Lambda$ interaction (Type 1) are shown in Fig.~\ref{fig:frac2}. 
The onset density of $\Lambda$ hyperons with the TBF is 0.59 fm$^{-3}$, which is higher than that without the TBF (0.42 fm$^{-3}$) due to the repulsion of the TBF. 
In addition, $\Sigma^-$ hyperons do not appear when the TBF is taken into account. 
In fact, the onset density of $\Sigma^-$ hyperons with the TBF is 1.50 fm$^{-3}$, which is higher than the critical density $n_{\mathrm{c}} = 1.13$ fm$^{-3}$. 
In other words, the TBF stiffens the EOS, which shifts the onset densities of hyperons to a higher density region.  
Furthermore, even if the $\Lambda$ mixing occurs at high densities, the EOS remains stiff because of the repulsive TBF.  
Owing to this repulsive effect, the present EOS with the TBF is sufficiently stiff to be consistent with the observational data. 
Here we note that, even with the TBF, the onset density of $\Lambda$ hyperons is insensitive to the odd-state $\Lambda \Lambda$ interaction, as for the case without the TBF shown in Fig.~\ref{fig:frac1}.

\section{Conclusions}
In this study, we have constructed the EOS of nuclear matter containing $\Lambda$ and $\Sigma^-$ hyperons by the cluster variational method. 
For the nucleon interactions, we employed the realistic AV18 two-body potential and UIX three-body potential. 
For the $\Lambda N$ interaction and the $\Lambda \Lambda$ even-state interaction, we employed the central three-range Gaussian potentials that are determined by reproducing the experimental data on single- and double-$\Lambda$ hypernuclei. 
Since there is no experimental data providing information on the odd-state $\Lambda \Lambda$ interaction, we constructed four models for it and investigated its influence on the structure of NSs.  
To this end, we employed the simple $\Sigma^- N$ interaction, which is determined so as to reproduce the experimental single-particle potential of the $\Sigma^-$ hyperons in symmetric nuclear matter at the saturation density.  
Starting from the Hamiltonian composed of these bare hyperon interactions, we calculated the energies of hyperonic nuclear matter for various particle fractions and apply the EOSs thus obtained to calculations of the structure of NSs. 

Owing to the hyperon mixing, the EOSs of NS matter with hyperons obtained by the variational method become softer than the EOS of pure nucleon matter.  
Correspondingly, the maximum mass of NSs with hyperons are less than that without hyperons. 
It is found that the maximum mass of NSs with the most repulsive $\Lambda \Lambda$ interaction is 1.62 $M_\odot$, whereas that with the most attractive $\Lambda \Lambda$ interaction is 1.48 $M_\odot$.  
Thus, the repulsion in the odd-state $\Lambda \Lambda$ interaction increases the maximum mass of NSs by about 9\%. 
In addition, an interesting result is that the onset density of $\Sigma^-$ hyperons in NS matter depends strongly on the odd-state $\Lambda \Lambda$ interaction, whereas that of $\Lambda$ hyperons is insensitive to this interaction.  
To our knowledge, these are the first results that describe how the odd-state $\Lambda \Lambda$ interaction affects the structure of NSs. 

Though the maximum mass of NSs increases because of the odd-state $\Lambda \Lambda$ repulsive interaction, it remains less than that given by the recent observational data on PSRs J1614-2230 and J0348+0432: 
One missing ingredient might be the repulsive TBF.  
Therefore, in this study, we introduce the universal TBF proposed in Refs.~\cite{MPP1, MPP2}. 
The obtained maximum mass of NSs with TBF is about 2.14 $M_{\odot}$, which is consistent with the masses of PSRs J1614-2230 and J0348+0432. 

Therefore, it is fascinating to study the influence of the TBF on the NS structure more in detail. 
Before those studies, however, we must clarify the properties of two-body $YN$ and $YY$ interactions, which also play important role in the structure of hypernuclei. 
In particular, we investigated, in this study, an important role of the odd-state $\Lambda \Lambda$ interaction in the NS structures. 
Thus, it is desirable to determine the odd-state $\Lambda \Lambda$ interaction by the E07 experiment at J-PARC.  
Furthermore, it is also interesting to calculate the excited states of double-$\Lambda$ hypernuclei, where the odd-state $\Lambda \Lambda$ interaction is important because one of the two $\Lambda$s occupies the $p$-wave state.  
Those investigations would provide more useful informations about the properties of NSs and $\Lambda$ hypernuclei.  
Moreover, mixing of other hyperons such as the $\Xi^-$ hyperon, which was not taken into account in this study, would be an important ingredient in the study of the NS structure.  
In particular, the attractive feature in the $\Xi^- N$ interaction was suggested recently \cite{XiN}.  
Therefore, it is also an important future problem to extend the present study so as to take into account mixing of other hyperons such as $\Xi^-$ hyperon.

\begin{acknowledgments}
H.~T. expresses thanks to Dr.~Y.~Funaki, Dr.~M.~Isaka, and Dr.~N.~Yamanaka for helpful discussions and comments. 
Early stages of this work were performed by Mr.~A.~Konno.  
The numerical computations in this work were carried out on SR16000 at the High Energy Accelerator Research Organization (KEK). 
This work is supported by JSPS (No. 23224006) and RIKEN iTHES Project. 
\end{acknowledgments}

\appendix
\section{Single-particle potential in the two-body cluster approximation} 
In this appendix, we present the explicit expression for the single-particle potential based on the cluster expansion.  
The cluster variational method was proposed by Iwamoto and Yamada \cite{IY} for fermion systems with state-independent two-body central forces. 
For neutron matter with two-body spin-dependent central interactions, the corresponding cluster expansion is summarized in the Appendix of Ref.~\cite{TY}.  
In the latter case, the single-particle energy $\epsilon_i$ is expressed in the two-body cluster approximation as 
\begin{eqnarray}
\epsilon_i &=& \int \varphi_i^\ast (x) H_i (x) \varphi_i (x) dx \nonumber \\
&+& \textstyle\sum\limits_{j} \int \int [\varphi_i^\ast (x_1) \varphi_j^\ast (x_2) - \varphi_j^\ast (x_1) \varphi_i^\ast (x_2)]   \nonumber   \\
&& \times f_{ij}^{\ast} H_{ij}(x_1, x_2) f_{ij} \varphi_i (x_1) \varphi_j (x_2)  dx_1 dx_2.  
\label{eq:ei}
\end{eqnarray}
Here, the explicit expressions of $H_i (x)$ and $H_{ij} (x_1, x_2)$ are given in Eqs.~(A$\cdot$2a) and (A$\cdot$2b) of Ref.~\cite{TY}, respectively. 

Extending the above expression for $\epsilon_i$ to hyperonic nuclear matter, the single-particle potential of a hyperon $U_Y (k)$ ($Y$ = $\Lambda$ or $\Sigma^-$) used in this study is given as a function of the wave number $k$ as follows: 
\begin{widetext}
\begin{eqnarray}
U_Y (k) &=& 2\pi n_{\mathrm{B}}{\textstyle\sum\limits_{b' = \mathrm{n}, \mathrm{p}, \Lambda, \Sigma^-}}\omega_{Yb'}{\textstyle\sum\limits_{p, s}} 
\int_0^{\infty}\Bigg[\big[f_{\mathrm{C}ps}^{\mu = Yb'}(r)\big]^2V_{\mathrm{C}ps}^{\mu = Yb'}(r) 
+ \frac{\hbar^2}{m_{\mu= Yb'}}\bigg[\frac{df_{\mathrm{C}ps}^{\mu = Yb'}(r)}{dr}\bigg]^2\Bigg]    \nonumber   \\ 
&& \times \frac{2s+1}{4} x_{b'} \bigg\{1 + \epsilon_p\bigg[3\frac{j_1(\xi_{b'}r)}{\xi_{b'}r}\bigg] j_0\big(\frac{m_{\mu = Yb'}}{m_Y}kr\big) \bigg\}r^2dr, 
\label{eq:sp}
\end{eqnarray}
\end{widetext}
where $\omega_{Yb'} = 1$ for $Y = b'$ or $\omega_{Yb'} = 1/2$ for $Y \neq b'$. 
For the central correlation functions $f_{\mathrm{C}ps}^{\mu}(r_{ij})$ on the right-hand side of Eq.~(\ref{eq:sp}), we employ the solutions of the Euler-Lagrange equations derived from Eq.~(\ref{eq:E2}). 
The single-particle potential $U_Y^0$ of hyperons in pure nucleon matter is then obtained with $U_Y^0 = U_Y (k = 0)$. 

\end{document}